
\voffset-1cm
\input amstex
\documentstyle{amsppt}

\topmatter
\title Enumeration of rational curves via torus actions
\endtitle
\author Maxim Kontsevich
\endauthor
\affil Max-Planck-Institut f\"ur Mathematik, Bonn
\\
and University of California, Berkeley
\endaffil
\address Dept. of Mathematics, University of California, Berkeley, CA 94720,
USA
\endaddress
\email maxim\@math.berkeley.edu
\endemail
\endtopmatter

\document

\NoBlackBoxes
\magnification=1200
\define\Q{\bold Q}
\define\C{\bold C}
\redefine\P{\bold P}
\redefine\O{\Cal O}
\define\E{\Cal E}
\define\F{\Cal F}
\define\T{\bold T}
\define\M{\overline{\Cal M}}
\define\eps{\epsilon}
\redefine\l{\lambda}
\define\Z{\bold Z}

 \head Introduction
\endhead

This paper contains an attempt to formulate rigorously and to check
predictions in enumerative geometry of curves following from Mirror Symmetry.

In a sense, we almost solved both problems.
There are still certain gaps in the foundations. Nevertheless, we obtain
``closed'' formulas for generating functions in the topological sigma-model for
a wide class of manifolds, covering many Calabi-Yau and Fano varieties. We
reduced Mirror Symmetry in a basic example  to a certain complicated but
explicit identity.
 We have made several computer checks. All results were as expected.  In
particular, we computed the ``physical'' number of rational curves of degree
$4$ on a quintic $3$-fold (during 5 minutes on Sun), which was out of reach of
previous algebro-geometric methods.

The text consists of 5 parts.
 The first part contains the definition of stable maps used through all the
paper. We establish several basic properties of  moduli spaces of stable maps.
 Also, we give an outline of a  construction of Gromov-Witten invariants
 for all algebraic projective or closed symplectic manifolds.
 For the reader who is interested mainly in computations it is enough to look
through 1.1 and to the statements of theorems occuring in 1.3.1-1.3.2.

In section 2 we describe a few examples of counting problems in the enumerative
geometry of curves. One of the examples is rational curves on quintics. We give
a simple algebro-geometric definition for the number of curves without assuming
the validity of the Clemens conjecture or using symplectic methods.

The main body of computations is contained in section 3.
  Our strategy here is quite standard: we reduce our problems to questions
concerning Chern classes on a space of rational curves lying in projective
spaces (A.~Altman - S.~Kleiman, S.~Katz), and then use Bott's residue formula
for the action of the group of diagonal matrices (G.~Ellingsrud and S.~A.~Str\o
mme).  As a result we get in all our examples  certain sums over trees.

In section 4 we develop a general scheme for summation over trees.
 By Feynman rules we know that such a sum should be equal to the critical value
of some functional. Using a trick we obtain an equivalent functional which is a
 quadratic polynomial in infinitely many variables with coefficients depending
on a finite number of variables. Thus, all our counting problems are reduced to
the inversion of  certain explicit square matrices with coefficients of
hypergeometric kind. This last step we were not able to accomplish.
Presumably, there is here a hidden structure of an integrable system and Sato's
grassmannians.

 In section 5 we describe extensions of our computation scheme to other
enumerative problems, including Calabi-Yau and Fano complete intersections (of
arbitrary dimension) in projective spaces, toric varieties and generalized flag
varieties.

\head 1. Stable maps.
\endhead

\subheading{1.1. Definition}

Let $V$ be a scheme of finite type over a field (or a smooth scheme, or a
complex manifold, or an almost complex manifold).

\proclaim{Definition}
A stable map is a structure $(C;x_1,\dots,x_k;f)$ consisting of a
 connected compact reduced curve $C$ with $k\ge 0$ pairwise distinct marked
non-singular points $x_i$ and at most ordinary double singular points, and a
map $f:C@>>> V$ having no non-trivial first order infinitesimal automorphisms,
identical on $V$ and $x_1,\dots,x_k$ (stability).
\endproclaim

The condition of stability means that every irreducible component
of $C$ of genus $0$ (resp. $1$) which maps to a point must have at least $3$
(resp. $1$) special (i.e. marked or singular) points on its normalization.
Also, it means that the automorphism group of $(C;x_1,\dots,x_k;f)$ is finite.

For a curve $C$ with at most ordinary double singular points its arithmetic
genus $p_a(C):=dim\,\,H^1(C,\Cal O)$ can be computed
 from the formula
$$2-2p_a(C)=\chi(C\setminus C ^{sing})\,\,.$$

Let $\beta\in H_2(V,\Z)$ be a homology class. (In the algebro-geometric
situation $\beta$ should be an element of the group of $1$-dimensional cycles
modulo homological equivalence).

\proclaim{Notation} $\M_{g,k}(V,\beta)$ denotes the moduli stack
of stable maps to $V$ of curves of arithmetic genus $g\ge 0$ with $k\ge 0$
marked points such that $f_*[C]=\beta$.
\endproclaim

More precisely, in the algebro-geometric setting one can define a family of
stable maps to $V$ as a flat proper morphism $\Cal C@>>>S$
 to a scheme $S$ of finite type over the ground field and a map
$f:\Cal C@>>> V$ such that its restriction to each geometric fiber of $\Cal C$
over $S$ is a stable map.

In the setting of almost complex manifolds we consider $\M_{g,k}(V,\beta)$ as a
set of equivalence classes of stable maps endowed with a natural topology (see
[P]) and an orbispace structure (see the next subsection).

\it Remark\rm. For many reasons one has to consider curves not in a fixed
manifold $V$ but in manifolds $V_{\lambda}$ varying in families. We have not
developed the corresponding formalism yet.
  In subsection 5.4 we describe a simple example of algebraic  $K3$-surfaces
which shows the necessity of families.
 It is also clear from our example that one can consider non-compact $V$ as
well.

\subheading{1.2. Orbispaces}

The notion of orbispace introduced here is  a topological counterpart
of
\roster
\item algebraic stacks (from algebraic geometry), and
\item orbifolds, or V-manifolds (from differential topology).
\endroster

We define an orbispace as a small topological category $C$ (i.e. a category for
which $\text{ Ob}\,C$ and $\text{ Mor}\,C$ carry topological structures)
satisfying the following axioms.

A.1. $C$ is a groupoid (every morphism is invertible).

A.2. For each $X,Y\in \text{ Ob}\,C$ the set of morphisms
  $Mor_C(X,Y)$ is finite.

A.3. The two maps from $\text{ Mor}\,C$ to $\text{ Ob}\,C$, assigning to a
morphism its source and its target respectively,
 are locally homeomorphisms (\'etale maps).

Functors between orbispaces which are continous, locally homeomorphisms and
induce an equivalence of categories we call equivalences between orbispaces.

The set $|S|$ of equivalence classes of objects of $C$ has a natural induced
topology. We can associate with each element $[X]\in |S|$ an equivalence class
(modulo interior automorphisms) of finite groups, $Aut(X)$.

\subheading{1.3. Properties of moduli spaces of stable maps}

The notion of a stable map is a mixture of the notion of a \it stable curve \rm
from algebraic geometry and of the notion of a  \it cusp-curve \rm
 from symplectic topology.
  By definition from [P], a cusp-curve is a holomorphic map $\widetilde f$ from
a compact (not necessarily connected) smooth complex curve $\widetilde C$ to an
almost-complex manifold $V$ and a finite collection $\Cal S$ of
non-intersecting $2$-element subsets of $\widetilde C$ such that, for each
$S\in \Cal S$,
 its image $\widetilde f(S)$ is a $1$-element set.
 By glueing points from pairs $S\in \Cal S$, we obtain a curve $C$ with at most
ordinary double singular points and a map $f:C@>>>V$. P.~Pansu claimed in [P]
that if $V$ is compact and endowed with a riemannian metric, then the space of
equivalence classes
 of cusp-curves of bounded genus and area is compact and Hausdorff.
  His claim is wrong, exactly because the condition of stability on components
which are mapped to a point was forgotten!
  It seems that, after appropriate corrections, the proof from [P] shows that
the moduli space of stable maps of bounded genus and area is compact and
Hausdorff.

Recall that in symplectic topology one considers usually
 almost-complex structures on symplectic manifolds compatible in an evident
sense with the symplectic form. Such a structure defines
 a riemannian metric on the underlying manifold, and the riemannian area of
each
 holomorphic curve coincides with its symplectic area. The latter is a pure
homological invariant. Hence $\M_{g,k}(V,\beta)$ is compact  and Hausdorff in
such a situation.

In  the next  subsection, we prove analogous properties of $\M_{g,k}(V,\beta)$
in the algebro-geometric setting.

In 1.3.2, we describe a situation in which the moduli space of stable maps is
smooth (as a stack).

\subheading{1.3.1. Algebraicity and properness}

\proclaim{Theorem} Let $V$ be a projective scheme of finite type over a field.
Then
 $\M_{g,k}(V,\beta)$ is an algebraic proper stack of finite type.
\endproclaim

The proof uses results from [DM]. We refer to [DM] for definitions  concerning
properties of stacks, and for other technical details as well.

We want to realize $\M_{g,n}(V,\beta)$ as a quotient stack of a scheme of
finite type modulo an \'etale equivalence relation.

  From the boundedness of the Hilbert scheme of $1$-dimensional subschemes of
$V$ it follows that for $(C;x_*;f)$ with fixed $p_a(C)$ and $\beta:=f_*[C]$,
the number of singular points on $C$ and the number of irreducible components
of $C$ are bounded.

In the next step, we will realize $\M_{g,n}(V,\beta)$ as a quotient space of a
space of maps of stable curves into $V$. For this we can choose a finite
collection of hypersurfaces $D_i$ in $V$ such that each non-stable component of
any curve $C$ from $\M_{g,n}(V,\beta)$
 intersects transversally  some of $D_i$ at least at three non-special points.
Then we can consider such intersection points as  new marked points on $C$.
Finally, we can glue a fixed smooth curve of genus bigger than $0$ with one
marked point to each marked point on $C$ obtaining a stable curve of  a bounded
genus. We map each glued  component into a point of $V$.

This way our moduli space is realized locally in the \'etale topology as a
closed subspace of  the space of maps from stable curves to $V$ with a fixed
image of the fundamental class. Such a space can be realized inside the Hilbert
scheme of $1$-dimensional subschemes  of the product of the universal curve
times $V$ via  graphs of maps.
 Thus,   $\M_{g,n}(V,\beta)$ is an algebraic stack of finite type.

Separatedness and properness of moduli of stable maps follow from
corresponding properties of $V$. Recall that the property of properness implies
separatedness by definition.

 If we have a family of stable maps  $\Cal C/K,\,\,f:\Cal C@>>>V$ over a
discrete valuation field $K$ with the ring of integers $\Cal O_K$, then there
exists a finite extension $L/K$ and a family of stable curves over $\Cal O_L$
extending the pull-back of $\Cal C/K$. First of all, we can construct a proper
two-dimensional scheme $S$ over $Spec (\Cal O_K)$ which maps to  $V$ by taking
the closure of the graph of $f$ into the product of $V$ and an arbitrary model
of $\Cal C$ over $\Cal O_K$. It follows from well-known facts about
degenerations of curves that there exists such an extension $L/K$ and a curve
$\Cal C'$ over $\Cal O_L$ which maps to $V$ with the property that the
geometric fiber of $\Cal C'$ over the closed point of $Spec(\Cal O_L)$ is a
connected reduced curve with pairwise distinct marked non-singular points and
at most ordinary double singular points. We can contract consecutively
non-stable components of this geometric fiber and obtain a stable map. This
proves the  existence part of the valuative criterion of properness.

  Moreover, in such a situation the components which we contract  all have
genus zero and form a subforest in the degeneration graph of the curve. One can
see easily that this subforest does not depend on the order in which we
contract components. From this uniqueness and separatedness of $V$ one can
conclude that the moduli stack of stable maps is separated. Hence, we have also
the uniqueness part of the valuative criterion of properness. \qed

\subheading{1.3.2. Smoothness}

\proclaim{Theorem} Let $V$ be a smooth proper scheme of finite type over a
field which is convex in the sense of [KM]. Then the stack
$\M_{0,k}(V,\beta)$ is smooth, and the complement of the open subset
$\M_{0,k}^0(V,\beta)$ consisting of smooth curves is a divisor with normal
crossings.\endproclaim

Recall that convex manifolds $V$ (definition 2.4.2 in [KM])
 are defined as projective manifolds with vanishing $H^1(C,f^*\Cal T_V)$ for
any stable map of genus zero. It is enough to check this only for smooth
curves. At the moment we know only one group of examples, namely, homogeneous
projective varieties. In sections 2-4   we will consider only projective
spaces.

Stable maps have the following important property.
  Let us consider  a flat proper morphism $\Cal C@>>>S$
 , of relative dimension $1$ to a scheme $S$ of finite type over the ground
field, sections $x_i,\,\,\,i=1,\dots, k$ of $\Cal C$ over $S$, and a map
$f:\Cal C@>>> V$. We claim that the set of points $p$ of $S$ such that the
restriction of $f$ to the geometric fiber of $\Cal C$ over $p$ is a stable map,
is an open subset of $S$. Hence, the deformation theory of a given stable map
is equivalent to the deformation theory   of it as  a map from a compact
(non-fixed) curve to $V$.

First of all, the deformations of $1$-st order of a stable map
 $(C;x_1,\dots,x_k;f)$ which do not change the structure of the singularities
of $C$ are given by
  $$\bold H^1(C,\Cal T'_{C}@>>>f^*\Cal T_V)\,,$$
where $\Cal T'_{C}$ denotes the sheaf of vector fields on $C$
 vanishing at the points $x_i,\,\,i=1,\dots,k$. We put $\Cal T'_{C}$ in  degree
$0$ and $ f^*\Cal T_V$ in  degree $1$. The hyper-cohomology group in degree $0$
vanishes by the stability condition.

Denote by $T$  the tangent space to  $\M_{g,k}(V,\beta)$ at the point
$(C;x_1,\dots,x_k;f)$.
 One can show that we have the following exact sequence:
$$0@>>>\bold H^1(C,\Cal F^{\bullet})@>>>T@>>>\bigoplus_
{y\in C^{sing}} T_y^1C\otimes T_y^2C@>>>\bold H^2(C,\Cal F^{\bullet})\,\,\,,$$
 where the fourth term comes from the deformations of $C$ resolving double
points $y$. The tangent spaces  to the two branches of $C$ at $y$ are denoted
by $T_y^1C$ and $T_y^2C$ (in arbitrary order); $\Cal F^{\bullet}$ denotes the
complex of sheaves of length $2$ used above.

We are ready now for the proof of the smoothness criterium.
 For an arbitrary map $f$ from a curve $C$ of arithmetic genus zero to a convex
$V$ we have
 $H^1(C,f^*\Cal T_V)=0$.
  Hence $\bold H^2(C,\Cal F^{\bullet})=0$, and the dimension of the tangent
space
 to $\M_{0,k}(V,\beta)$ is constant. One can elaborate the argument above for
maps parametrized by spectra of Artin algebras and  show that there is no
obstructions for the deformation theory.
  Also, we see that maps from singular curves form a divisor with normal
crossings.
\qed

{}From the proven properties of $\M_{0,k}(V,\beta)$ one can easily deduce the
tree level system of Gromov-Witten invariants on convex varieties (see [KM]).

\subheading{1.4. The structure of an intersection of manifolds}

The last theorem shows that the moduli space of stable maps to $V$ inherits
 the property of smoothness of $V$ in some cases. Here we are trying to define
 for \it all \rm smooth $V$ a certain structure on $\M_{g,k}(V,\beta)$
 which permits us to construct an analogue of the fundamental class.
 We will do it in the setting of almost-complex real-analytic manifolds and
describe in 1.4.2 the situation in algebraic geometry.

Let $Y_1,Y_2$ be two submanifolds in a manifold $X$ (manifolds are
real-analytic, or complex, or algebraic). The intersection $Z:=Y_1\cap Y_2$
 in general is not smooth. Nevertheless, we define its
 ``virtual tangent bundle'' $[\Cal T_Z]^{virt}\in K^0(Z)$
  by the formula
 $$[\Cal T_Z]^{virt}:=[\Cal T_{Y_1}]_{|Z}+[\Cal T_{Y_2}]_{|Z}-
[\Cal T_X]_{|Z}\,\,.$$
Also, if $X,Y_1,Y_2$ are oriented then there is a canonical
 ``virtual fundamental class'' $[Z]^{virt}$ with values in homology with closed
support
$$H_d^{closed}(Z):=\widetilde H_d(\widetilde Z,\Z)\,\,.$$
 Here $\widetilde Z$ denotes the one-point compactification of $Z$ and
$\widetilde H_d$ denotes the $d$-th reduced homology group of a punctured
space. The number $d=dim(Y_1)+dim(Y_2)-dim(X)$ is the virtual dimension of $Z$.
 More precisely, one can construct a fundamental class in the   complex
bordism group  with closed support $\Omega^{closed}_d(Z)$, defined analogously,
when $X,Y_1,Y_2$ are almost-complex. The idea is obvious: $Z$ is homotopy
equivalent to a sufficiently small tubular neighborhood $UZ$ of $Z$ in $X$. In
$UZ$ one can perturb generically $Y_1,Y_2$ and obtain a transversal
intersection. Notice that in the smooth situation $Z$ can have a pathological
topology and can be not homotopy equivalent to any  tubular neighborhood.  It
is plausible that
 one can define a cobordism analogue of Borel-Moore homology
 and extend intersection theory to the smooth case.

The singular space $Z$ can have several representations as an intersection of
germs at $Z$ of manifolds containing $Z$. For example, we can multiply
$X,Y_1,Y_2$
 locally by $X',Y'_1,Y'_2$ where $Y'_1$ intersects $Y'_2$ transversally in one
point. Globally one can pass from $X,Y_1,Y_2$ to the total spaces of vector
bundles $\Cal E^X,\Cal E^{Y_1},\Cal E^{Y_2}$ on corresponding spaces endowed
with embeddings
$$\Cal E^{Y_i}\hookrightarrow\Cal E^X_{|Y_i}\,,\,\,\,\,\Cal  E^X_{|Z}\simeq\Cal
E^{Y_1}_{|Z}\oplus
\Cal E^{Y_2}_{|Z}\,\,.$$
Such pairs of representations we call stably equivalent.

If $Z$ is an intersection of several submanifolds $Y_i,\,1\le i\le n$ in $X$,
then one can represent $Z$ as an intersection of two submanifolds:
 $$Z\simeq (Y_1\times \dots \times Y_n)\cap \text{ diagonal in }X^n\,\,.$$
Also, if we have two maps of manifolds
 $Y_i@>f_i>>X,\,\,i=1,2$ then the fiber product $Z:=Y_1\times_X Y_2$ carries a
structure of an intersection of two manifolds. It follows from the
identification of $Z$ with the intersection in $Y_1\times Y_2\times X$ of the
graphs of $f_i$ multiplied by $Y_{3-i}$.

Let the space $Z$ carry a system of representations of open subsets
 $U_i$ of $Z$ as intersection of  manifolds endowed with a system of stable
equivalence between models for $U_i\cap U_j$ arising from $U_i$ and $U_j$. Such
a system should be associative up to a homotopy, homotopies between homotopies
etc. Then we expect that $Z$  has a global virtual tangent bundle and a virtual
fundamental class.
 In a sense, all this should be a non-linear analogue of an element of $K^0(Z)$
represented locally as a formal difference of two vector bundles.

Let us return to the moduli space of stable maps.
We claim that it has  a canonical structure of an orbifoldic version  of an
intersection of almost-complex manifolds.

First of all, near each
 point $(C;x_1,\dots,x_k;f)$ we will represent $\M_{g,k}(V,\beta)$ as an
intersection of several infinite-dimensional Frechet submanifolds (forgetting
temporarily the presence of the finite group of automorphisms of the stable map
$(C;x_1,\dots,x_k;f)$). Let us choose several closed non-intersecting simple
parametrized loops $L_i$ on the surface $C$ which divide it into pieces $C_j$
each of which is either a smooth surface with a non-empty boundary and no
marked points, or a smooth disc with one marked point in the interior of the
disc, or  two discs with glued centers.

We consider as the first approximation to the ambient manifold $X$,  the space
$X'$ of smooth maps  from $\coprod L_i$ into $V$ which are sufficiently close
to $f_{|\coprod L_i}$. This space is an infinite-dimensional almost complex
manifold with the complex structure on the tangent bundle induced pointwise
from the complex structure on $\Cal T_V$.

For each piece $C_j$ of the surface $C$ we introduce the space $Y'_j$
 consisting of pairs $(J',f')$, where $J'$ is a complex structure on  $C_j$
close to the initial one and $f'$ is a $J'$-holomorphic map  considered modulo
diffeomorphisms of $C_j$ close to the identity.
  For pieces $C_j$ which are two intersecting discs we add  small flat
deformations:
$$\{(x,y):\,x,y\in \C,\,\,xy=0,\,\,|x|+|y|\le 1\}$$
$$\text{ deform to }
  \{(x,y):\,x,y\in \C,\,\,xy=\eps,\,\,|x|+|y|\le 1\}\,,\,\,\,|\eps|\ll 1\,.$$

The spaces $Y'_j$ are almost-complex and they are mapped into $X'$ by passing
to the  restriction  of maps to  boundaries. Their fiber product over $X'$
consists of stable maps of curves close to the initial point endowed with
parametrized loops. One can pass to the quotient of all the picture modulo the
action of  the product over loops $L_i$  of the ``complexified diffeomorphism
group of a circle''(= replacing curves $L_i$ by close curves). This action is
free exactly because of the condition of stability. Finally, we can pass to the
case of two submanifolds, as we already explained.

Thus, we will get germs of Frechet manifolds $X$ and $Y_i$. The natural map of
tangent spaces
$$T_zY_1\oplus T_zY_2@>>>T_zX\,,\,\,\,z\in Z\,\,,$$
is Fredholm.
In the next subsection we develop a technique producing in such a situation
finite-dimensional models.

Globally, we can cover $Z=\M_{g,k}(V,\beta)$ by finitely many open sets: and on
each of them  we have an equivalence class of representations as intersections
of manifolds. It is almost clear
 that different representations on intersections of  open sets
 are equivalent modulo homotopy and higher homotopies between homotopies on
multiple intersections. Unfortunately, we do not know
 how to formulate all this  precisely.
We have tried to avoid choices and use a Dolbeault-type resolvent.
 The  space $Y_1$ in this case should be a space of real-analytic maps from
complex surfaces to $V$ satisfying the same condition of stability as before. A
natural candidate for $X$ will be the total space of a vector bundle on $X$
arising from the Cauchy-Riemann equation, $Y_2$ will be a section of $X$ as of
a vector bundle.  We  have  met an unpleasant difficulty
 in considering deformations resolving double points. Maybe, nevertheless, it
is possible to find a  version of the Dolbeault complex which forms   a complex
of infinite-dimensional vector  bundles over a neighborhood of $Z$ in $Y_1$.
Such a construction, if it exists, will give a really simple definition of the
virtual fundamental class of $Z$.

\subheading{1.4.1. Reduction to finite dimensional manifolds}

Suppose that we have Frechet manifolds $X$ and $Y_i$ with the  Fredholm
property as above and with $Z:=Y_1\cap Y_2$  compact.

We can choose smooth finite-dimensional sub-bundles $\Cal E_i\subset (\Cal
T_X)_{|Z},\,i=1,2\,\,$ such that
 $$\Cal E_i\cap (\Cal T_{Y_i})_{|Z}=0\,,\,\,\,\,
 (\Cal T_X)_{|Z}=\Cal E_1+\Cal E_2+(\Cal T_{Y_1})_{|Z}+(\Cal
T_{Y_2})_{|Z}\,\,\,.$$
We can prolong $\Cal E_i$ to  neigborhoods of $Z$ in $Y_i$.
Then we can choose  submanifolds $\widetilde Y_i$ in $X$ containing $Y_i$ such
that
$$(\Cal T_{\widetilde Y_i})_{|Y_i}=\Cal T_{Y_i}\oplus \Cal E_i\,\,.$$
The submanifolds $\widetilde Y_i\subset X$ intersect each other transversally
near $Z$. From now on we can forget about the ambient manifold $X$ and consider
only the system of $5$ manifolds and inclusions between them:
$$Y_1\hookrightarrow \widetilde Y_1\hookleftarrow\widetilde Y_1\cap\widetilde
Y_2\hookrightarrow \widetilde Y_2\hookleftarrow Y_2\,\,.$$

We can choose  sub-bundles $\Cal F_i$ of finite codimension in $(\Cal
T_{Y_i})_{|Z}$ such that
  $$\Cal F_i\cap(\Cal T_{\widetilde Y_{3-i}})_{|Z}=0\,\,.$$
Then we can choose a foliation in $Y_i$ with tangent spaces to fibers at points
from $Z$ equal to $\Cal F_i$. We can prolong these
 foliations to foliations of $\widetilde Y_i$. Near $Z$ these foliations are
tangent to fibers of smooth fibrations, due to the transversality
 condition above.  Passing to the spaces of fibers of these foliations in
$Y_i,\widetilde Y_i$ we obtain germs of finite-dimensional manifolds
$Y_i',\widetilde Y_i'$.
  We can take for  the middle term the same finite-dimensional manifold as
above, $\widetilde Y_1\cap\widetilde Y_2$, and get a new finite-dimensional
system of $5$ manifolds and inclusions. We can construct a new  ambient
manifold $X'$ in which $\widetilde Y_1'$  and $\widetilde Y_2'$ intersect
transversally  along
 $\widetilde Y_1'\cap\widetilde Y_2' \simeq\widetilde Y_1\cap\widetilde Y_2$.
 Thus,  $Z$ is realized as an intersection of finite-dimensional manifolds. One
can check that
 different procedures give  stably equivalent representations.

\subheading{1.4.2. Intersections of manifolds in algebraic geometry}

 If $Y_1,Y_2$ are submanifolds of an algebraic manifold $X$ then on $Z:=Y_1\cap
Y_2$ we can construct a structure of \it super-scheme\rm. This means that on
$Z$ we have a super-structure sheaf

 $$\Cal O_Z^*=\bigoplus_{n\le 0} \Cal O^n_Z$$
of $\Z_{\le 0}$-graded super-commutative rings such that $Z$ is a scheme with
respect to $\Cal O^0_Z$ and the $\Cal O^n_Z$ are coherent sheaves of $\Cal
O^0_Z$-modules.  The formula for the components of the higher structure sheaf
is
$$\Cal O^n_Z:=\, (i_Z)^* \,Tor_{-n}^{\Cal O_X}\left( (i_{Y_1})_*\Cal
O_{Y_1},(i_{Y_2})_*\Cal O_{Y_2}\right)\,\,,$$
where $i_?$ denotes the embedding map.
  Also we have a virtual tangent bundle in $K^0(Z)$ given by the same formula
as in the almost complex setting.

The structures ($\Cal O_Z^*,\,[\Cal T_Z]^{virt}$) do not change if we pass to
an equivalent representation of $Z$ as an intersection of two manifolds. We
call a pair of such structures a ``quasi-manifold''.

Our discussion leads to the prediction of the existence of the structure of a
quasi-manifold on $Z=\M_{g,k}(V,\beta)$
  defined in purely algebro-geometric terms.
  In fact, one can define $[\Cal T_Z]^{virt}$ as the direct image of the
deformation sheaf on the universal curve. Also, $\Cal O^0_Z$
 is the usual structure sheaf on the algebraic stack $Z$, and $\Cal O^1_Z$ is
equivalent  to the first obstruction sheaf. We are planning to write later more
about the definitions of higher structure sheaves and virtual tangent bundles,
arising ubiquitously in algebraic geometry. For example, various moduli spaces
and Hilbert schemes should carry canonical structures of quasi-manifolds.
  The idea of introducing higher structure sheaves on moduli spaces occured
implicitly quite a long time ago. It was recently spelled  out clearly in a
letter of P.~Deligne to H.~Esnault, together with a proposal to apply it  to
the algebro-geometric formulation of Mirror Symmetry.

We finish this subsection with a formula which produces a  virtual fundamental
class for a quasi-manifold $Z$ with $\Cal O^n_Z=0$
 for $n\ll 0$. Note that it is applicable to the quasi-manifold structure
arising on the intersection of two manifolds.

First of all, for each separated scheme $Z$ of finite type over a field
 and for any coherent sheaf $\Cal F$ on $Z$ a homological Chern class
$\tau(\Cal F)\in CH_*(Z)\otimes \Q$ is defined (see [BFM]).  Here $CH_*(Z)$
denotes the Chow group of cycles on $Z$ modulo rational equivalence, to be
regarded as an algebraic counterpart of $H_*^{closed}(Z(\C),\Z)$ for schemes
over $\C$.
For the definition of the virtual fundamental class, we will use, for the sake
of simplicity, a smooth ambient manifold $X$.
 We can prolong the virtual tangent bundle $[\Cal T_Z]^{virt}$
 to an element of $K^0(X)$ after replacing $X$ by a sufficiently small
neighborhood of $Z$. The formula for the virtual fundamental class is
$$[Z]^{virt}:=\left(\sum_k (-1)^k \tau(\Cal O^k_Z)\right)\cap td([\Cal
T_Z]^{virt})^{-1}\,\,.$$

One can see that for quasi-manifolds arising as an intersection of two
submanifolds this formula gives the same class as the usual intersection theory
of Baum-Fulton-MacPherson. For zero-dimensional $Z$ our formula is equivalent
to the Serre formula for multiplicities.

 Note that we have for quasi-manifolds a refined fundamental class with values
in $CH_*\otimes \Q[c_1,c_2,\dots]$ arising from the action of
 the Chern classes of $[\Cal T_Z]^{virt}$ on $[Z]^{virt}$. It can be considered
as an algebraic version of the fundamental class with values in complex
cobordism groups.

\subheading{1.5. Gromov-Witten invariants}

We have a compact moduli orbi-space of stable curves and a virtual fundamental
class of it in two situations:
\roster
\item $V$ is a smooth projective algebraic manifold over a field, or
\item $V$ is a compact real-analytic symplectic manifold endowed with a
compatible real-analytic almost-complex structure.
\endroster

The virtual fundamental class takes values in $CH_*\otimes \Q[c_1,c_2,\dots]$
and in $\Omega_*\otimes\Q$ respectively.

Suppose that $2-2g-k<0$ so that $\M_{g,k}$ exists.
 We have an evident map:
$$\Phi:\M_{g,k}(V,\beta)@>>>V^k\times\M_{g,k}\,,$$
$$(C;x_1,\dots,x_k;f)\mapsto \left(f(x_1),\dots,f(x_k);(\widetilde C,\tilde
x_1,\dots,\tilde x_k)\right)\,\,.$$
Here $(\widetilde C,\tilde x_1,\dots,\tilde x_k)$ is the stable curve with
marked points obtained from $(C,x_1,\dots,x_k)$ by consecutive contractions of
non-stable components.

The image under $\Phi$ of  $[\M_{g,k}(V,\beta)]^{virt}$ is a
 class in $V^k\times\M_{g,k}$ which leads to Gromov-Witten invariants of $V$
(see [KM]). It should not depend on the choice of  an almost-complex structure
in the symplectic case.

We expect that these classes satisfy all axioms postulated in [KM]. In fact,
the definition of stable maps was designed specially for this purpose.

Here we get a refinement of the  picture from [KM]: Gromov-Witten invariants
take their values not just in cohomology groups, but in complex cobordism
groups. Also, we have line bundles on $\M_{g,k}(V,\beta)$ with fibers equal to
$T_{x_i}C$, and we can take in account actions of their Chern classes on
$[\M_{g,k}(V,\beta)]^{virt}$. It is an essential additional data,
 because the $T_{x_i}C$ are not isomorphic to the pullbacks of analogous
bundles on $\M_{g,k}$. In the deformation formula 6.4.c from [KM],  one can use
$T_{x_i}C$
 instead of $T_{\tilde x_i}\widetilde C$.

\subheading{1.6. Comparison with other definitions}

It was proposed earlier several times that for the definition of the
topological sigma model (=Gromov-Witten invariants) in algebro-geometric terms
one should use the Hilbert scheme of $V$  and, possibly, modify it. Our moduli
space of stable maps does it in a sense. Its advantage is smoothness in the
case when $V$ is a generalized flag variety.
 Also, our definition gives the same moduli space for complex projective $V$
considered as an algebraic or as a symplectic manifold.

In symplectic geometry the most advanced construction was announced recently by
Y.~Ruan and G.~Tian in [RT]. They construct a part of the Gromov-Witten
invariants (essentially, genus zero invariants) in the case of semi-positive
symplectic manifolds. Their main idea is that in this case one can ignore
curves with singularities, because the dimension of the space of degenerate
curves is strictly less than the dimension of the space of smooth curves for
generic almost-complex structures. The advantage of the approach of [RT] is a
control on integrality of arising homology classes.
 In [RT] the  ``numbers of rational curves'' of a fixed homology class passing
through several submanifolds were defined in the case of the number of cycles
greater than or equal to $3$. Without this condition the ``number of curves''
should be fractional in the examples of quintic 3-fold (see the next section).

Our pre-definition should work for all symplectic manifolds and, presumably, in
the case of surfaces with boundaries, open a way to extend Floer's proof of the
Arnold conjecture to the case of non semi-positive symplectic manifolds.

As we already noticed, Gromov-Witten invariants should be defined    also for
families of not necessarily compact symplectic or algebraic manifolds. It is
not clear at the moment in which generality such a theory can be developed. For
example, we do not know whether families should  be flat or only smooth,
whether  the parameter space  should be smooth, etc.

The general scheme described in 1.4 can be applied in other  situations: moduli
 of vector bundles on algebraic curves and surfaces, moduli  of complex
structures on surfaces, moduli of vector bundles on Calabi-Yau 3-folds. The
common property of all such examples is that the natural complex whose $1$-st
cohomology group is equivalent to the tangent space to the appropriate moduli
space, has trivial cohomology in degrees greater than or equal to $3$. The main
 problem
 is to define  good compactifications in the other situations.

\head 2. Three examples.
\endhead

In this section and in the next one we
 will use  simplified notations:
$\M_{g,k}(\P^n,d[\P^1])$ will be denoted by
 $\M_{g,k}(\P^n,d)$, or simply by $\M(\P^n,d)$ if $g=k=0$.

\subheading{2.1. Rational curves on $\P^2$}

The dimension of the space $\M_{0,k}(\P^2,d)$ is equal to $3d-1+k$.
 For $k=3d-1$ it coincides with the dimension of $(\P^2)^k$.
 Hence
the number $P_d$ of rational curves on $\P^2$ of degree
$d\ge 1$ passing through generic $k=3d-1$ points is finite and equal to the
degree of the map $$\phi:\M_{0,k}(\P^2,d)@>>>(\P^2)^k,\qquad
 \phi(C; {x_1,\dots,x_k};f)=(f(x_1),\dots,f(x_k))\,.$$

 We can rewrite it as the integral:

$$P_d\,\,=\int\limits_{\M_{0,3d-1}(\P^2,d)}\prod_{i=1}^{3d-1}\phi^*(c_1(\O(1)_i)^2)\,\,,$$
where $\O(1)_i$ denotes the pullback of the line bundle $\O(1)$ from the $i$-th
factor $\P^2$ of $(\P^2)^{3d-1}$.

These numbers are known from the recursion relations following from
the associativity equations (see [KM]).
 The first few values of $P_d$ are:
$$\matrix
d & 1 & 2 & 3 & 4 &5\\
P_d& 1 & 1& 12& 620&87304
\endmatrix$$

Our proof of the associativity relations is based (following Witten [W]) on a
study of the boundary divisors of the moduli spaces of stable maps.
 Here we want to compute this number of curves directly.

\subheading{2.2. Rational curves on quintics}

 Let a smooth quintic 3-fold $V$ be given by an equation
$Q(x_1,\dots,x_5)=0$ in homogeneous coordinates in $\P^4$.
 The polynomial $Q$ of degree $5$ can be considered as a section
 of the line bundle $\O(5)$ on $\P^4$.

The orbispace $\M_{0,0}(V,d[\P^1])$ is a subspace of $\M(\P^4,d)$.
  Let $\E_d$ be a coherent sheaf on $\M(\P^4,d)$
 equal to the direct image under the forgetful map
 $\M_{0,1}(\P^4,d)@>>>\M_{0,0}(\P^4,d)=\M(\P^4,d)$ of $\phi^*(\O(5))$.
  Here again $\phi(C;x_1;f)=f(x_1)\in \P^4$.
 The sheaf $\Cal E_d$ is actually a vector bundle:
  for any stable map $f:C@>>>\P^4$ from a curve of arithmetic genus zero
$H^1(C,f^*(\O(5)))=0$, because the line bundle $\O(5)$ is generated by its
global sections.

The section $Q$ of $\O(5)$ defines sections $\widetilde{Q}_d$ of
$\E_d$ for all $d$. It is clear that $\M(V,d[\P^1])$ coincides with the scheme
of zeroes of $\widetilde{Q}_d$.
 We claim that this identification is compatible with the structure of an
intersection of manifolds.

The orbifold $\M(\P^4,d)$ has dimension $5d+1$, the same as the
rank of $\E_d$. Hence, we get the algebro-geometric definition
 of the ``number of rational curves on a quintic''. It should be equal to the
integral of the Euler class of $\E_d$:

$$N_d:=\int\limits_{\M(\P^4,d)}c_{5d+1}(\E_d)\,.$$

The numbers $N_d$ are not integers, because we use orbifolds.
The table of the first few $N_d$ is the following:

$$\matrix d& 1 &2 &3 &4\\
N_d &
\frac{2875}{1}&\frac{4876875}{8}&\frac{8564575000}{27}&\frac{15517926796875}{64}\endmatrix$$

The numbers $N_d$ are related with integer numbers $N_d^0$
 by the following formula:
$$N_d=\sum_{k:k|d}k^{-3}N^0_{d/k}\,\,.$$

$N^0_d$ is the number of geometric (unparametrized) rational curves
on $V$ with generically perturbed almost complex structure.

Mirror Symmetry (see [Y]) gives the following description of the sequence
$N_d$:

Let us introduce a function defined in a domain $\{t:\,Re(t)\ll 0,\,|Im(t)|<
\pi\}$ in the complex affine line
 $\C$:
$$F(t):=\frac{5}{6}t^3+\sum_{d\ge 1} N_d e^{dt}\,\,.$$
 We denote by $G(q_1,q_2)$ the corresponding function of homogeneity
 degree $2$ in a domain of the vector space $\C^2$:
 $$G(q_1,q_2)=F(q_1/q_2)\,q_2^2\,\,.$$
  The function $G$ generates a Lagrangean cone $\Cal L$ in the symplectic
vector space $\C^4$:
$$\Cal L\,:=\{(p_1,p_2,q_1,q_2):\,p_i=\partial G/\partial q_i\}\,\,.$$

On the other hand,
 $$I(z)=\sum_{n\ge 0} \frac{(5n)!}{(n!)^5}z^n$$
 is one of the periods of the one-dimensional variation of Hodge structures
$\Cal H_z $ with Hodge numbers $h^{0,3}=h^{1,2}=h^{2,1}=h^{3,0}=1$
 arising from a mirror family of Calabi-Yau $3$-folds.
  The Poincar\'e pairing defines a covariantly constant symplectic structure on
$4$-dimensional vector bundle $\Cal H$ with a flat  Gauss-Manin connection.  We
can trivialize the flat bundle $\Cal H$
 in the domain $\{z: |z|\ll 1,\,\,|Arg(z)|< \pi\}$. The union $\Cal U$ of the
$1$-dimensional terms of the Hodge filtration $F^3_z$ forms a Lagrangean cone
in $\C^4$.

Mirror Symmetry predicts that $\Cal L=\Cal U$.
 The same kind of correspondence is expected for other Calabi-Yau
 $3$-folds.

\subheading{2.3. Multiple coverings of rational curves on Calabi-Yau 3-folds}

Let $C_0\simeq\P^1$ be a  smooth rational curve in a complex 3-fold $V$ with
the normal bundle $\Cal T_V/\Cal T_{C_0}$ equivalent to $\O(-1)\oplus\O(-1)$.
In such a situation  $\M_{0,0}(V,d[C_0])$ has a connected component
$\M_{0,0}(C_0,d[C_0])$ consisting of  stable maps
 $C@>>>C_0$ of degree $d$. This component is isomorphic to $\M(\P^1,d)$ and has
dimension $2d-2$. The virtual dimension
 of this component is zero. The obstruction sheaf $\F_d$ is a vector bundle of
rank $2d-2$ with the fiber at each point $f:C@>>>C_0$
 equal to
$$H^1(C,f^*(\Cal T_V/\Cal T_{C_0}))\simeq\C^2\otimes H^1(C,f^*(\O(-1)))\,.$$
Our definition
 of the contribution of $\M_{C_0}$ will be the integral over it
of the Euler class of the obstruction sheaf:

$$M_d:=\int\limits_{\M(\P^1,d)} c_{2d-2}(\F_d)\,\,.$$

After Aspinwall and Morrison we expect that
$M_d=d^{-3}$. This was checked recently by Yu.~Manin (see [M]).

Actually, it is not clear why the computations from [AM]
 give the same answer as we get with the stable curves.
 In [AM] the authors consider the space of maps of rational curves with 3
marked
 points on it into $\P^1$ and compactify it by means of the Hilbert scheme of
1-dimensional subschemes in $\P^1\times \P^1$
 (they associate with a map $f:\P^1@>>>\P^1$ its graph).
Then they use the Euler class of a natural candidate for the obstruction bundle
and intersect it with a class of codimension $3$.
The answer which they get is $1$.
Of course, we can use $\M_{0,3}(\P^1,d)$ instead of $\M_{0,0}(\P^1,d)$ and
 modify the definition of ``the number of curves'' following the sample 2.1.
One can see easily that the result will be $d^3 M_d$.
  At the moment we do not know how to relate the  compactification from [AM]
and the moduli spaces of stable maps.

\head 3. Fixed points formulas.
\endhead

\subheading{3.1. The Bott fixed points formula}

 Let $X$ be a smooth compact complex projective manifold, and let $\Cal E$  be
a
holomorphic vector bundle on $X$. We suppose that a complex torus
 $\T\simeq\C^{\times}\times \dots\times \C^{\times}$ acts algebraically on
$(X,\Cal E)$.

Bott's formula reduces the computation of integrals of characteristic classes
of $\Cal E$ over $X$ to
  computations on the subspace of fixed points $X^{\T}$.
  This space is always a union of smooth subvarieties,  because the real
subgroup $\T^{real}=U(1)\times\dots\times U(1)$ is compact.
We denote connected components of
$X^{\T}$ by $X^{\gamma}$.

On each component $X^{\gamma}$ the vector bundle $\Cal E$
  splits into the direct sum of bundles $E^{\gamma,\l}$ over  characters
$\l:\T@>>>\C^{\times},\,\,\l\in\T^{\vee}\simeq\Z\oplus\dots\oplus\Z$. Also, the
normal bundle
 $\Cal N^{\gamma}=\Cal T_X/\Cal T_{X^{\gamma}}$ splits into the direct sum of
bundles $\Cal N^{\gamma,\l}, \,\,\l\in\T^{\vee}\setminus\{0\}$.

We add to $H^{even}(X,\Q)$ extra generators $e_i,\,\,i=1,\dots ,rk(\Cal E)$ of
degree $2$  obeying the relations
 $$\dsize\sum_{k\ge 0}c_k(\Cal E)=\prod_{i}(1+e_i)\,\,.$$ Analogously, we add
generators $e^{\gamma,\l}_i$
 and $n^{\gamma,\l}_i$ to $H^{ev}(X^{\gamma},\Q)$.

Let $P$ be a homogeneous symmetric polynomial (in a sufficiently large number
of variables) of degree $dim_{\C}(X)$. Bott's formula reads:

$$\int\limits_{X} P(e_i)=\sum_{\gamma}\int\limits_{X^{\gamma}}
 \frac{P(e^{\gamma,\l}_i+\l)}{\prod (n^{\gamma,\l}_i+\l)}\,\,.$$

Here the r.h.s. is considered as a rational function on $Lie(\T)$
 (each character $\l$ defines a linear form on $Lie(\T)$).
 In the numerator and the denominator we use all generators $e^{\gamma,\l}_i$
 and $n^{\gamma,\l}$ with fixed index $\gamma$.
An analogous  formula is valid for finite collections $(\Cal E^{(i)})_{i=1,N}$
of equivariant vector bundles and homogeneous polynomials $P$ in $N$ groups of
variables symmetric inside each group.

This formula is valid for orbifolds too, because the original proof
 in [B] uses only the language of differential forms and
 transfers immediately to the more general setting of orbifolds.

\subheading{3.2. Fixed points on moduli spaces of stable maps}

The action of the group $\T\simeq(\C^{\times})^{n+1}$ of diagonal matrices on
$\P^n$ induces
 an action of $\T$ on $\M_{g,k}(\P^n,d)$. We will describe the set of fixed
points in this subsection.

Denote by $p_i,\,\,i=1,\dots ,n+1$ the fixed points of $\T$ acting on $\P^n$.
The point $p_i$ is the projectivization of the $i$-th coordinate line in
$\C^{n+1}$. Also, denote by $l_{ij}=l_{ji},\,\,i\ne j$
 the line in $\P^n$ passing through $p_i$ and $p_j$.

Let a stable map
 $f:C@>>>\P^n$ represent a point of $\M_{g,k}(\P^n,d)^{\T}$. First of all, the
geometric image of $f$ should be invariant under the $\T$-action.
 One can see easily that this means that $f(C)$ is a union of lines $l_{ij}$.
Secondly, the images of all marked and singular points, as well as of
components of $C$ contracted by $f$,
 should be points $p_i$. Thirdly,
  each irreducible component $C^{\alpha}$ of $C$ which does not map to a point
has genus zero and maps onto one of the lines $l_{ij}$. In some homogeneous
coordinates it is given by

$$f(z_1:z_2)=(0:\,\,\dots\,\,:0:z_1^{d_{\alpha}}:0:\,\,\dots\,\,:0:z_2^{d_{\alpha}}:0:\,\,\dots\,\,:0)\,,\,\,\,{d_{\alpha}}\ge1\,\,.$$

We will associate  with each point
$(C;{x_1,\dots,x_k};f)\in\M_{g,k}(\P^n,d)^{\T}$
 a graph $\Gamma$. By \it graph \rm we mean a finite $1$-dimensional
$CW$-complex.
 The vertices $v\in Vert(\Gamma)$ correspond to the  connected components
$C_{v}$ of $f^{-1}({p_1,\dots,p_{n+1}})$. Note that each $C_v$ can be either a
point of $C$ or a non-empty union of irreducible components of $C$. The edges
$\alpha\in Edge(\Gamma)$ correspond to irreducible components $C^{\alpha}$ of
genus zero mapping to lines $l_{ij}$.
  We  endow $\Gamma$ with additional specifications:
 the vertices $v$ will be labeled by numbers $f_v$ from $1$ to $n+1$
 defined by the formula $f(C_v)=p_{f_v}$. The edges will be labeled by the
degrees ${d_{\alpha}}\in\bold N$. Also, we associate with each vertex $v\in
Vert(\Gamma)$ its interior genus $g_v$ (=arithmetic genus of the
$1$-dimensional part of
 $C_v\subset C$) and the set $S_v\subset\{1,\dots,k\}$ of indices of  marked
points lying on $C_v$.

Our claim is that connected components of $ \M_{g,k}(\P^n,d)^{\T}$ are
naturally labeled by equivalence classes of connected graphs $\Gamma$ with
specifications
 obeying the following conditions:
\roster
\item if  $\alpha\in Edge (\Gamma)$ connects vertices $v,u\in Vert(\Gamma)$
then $f_v\ne f_u\,\,$,
\item $1-\chi(\Gamma)+\dsize\sum_{v\in Vert(\Gamma)}g_v=g\,\,$,
\item $\dsize\sum_{\alpha\in Edge(\Gamma)}d_{\alpha}=d\,\,$,
\item $\{1,\dots,k\}=\dsize\coprod_{v\in Vert(\Gamma)} S_v\,\,$.
\endroster
Notice that from  condition (1) it follows that $\Gamma$ has no simple loops.

 Each component $\M_{g,k}(\P^n,d)^{\Gamma}$ is isomorphic to the quotient space
of the product of moduli spaces of stable curves over the set of vertices of
$\Gamma$ modulo the action of the automorphism group of $\Gamma$. We will
forget about $Aut(\Gamma)$ till 3.4.

\subheading{3.3. Contributions of connected components}

{}From now on we assume that our curves have arithmetic genus zero.
 The graphs $\Gamma$ in our description will be  trees and the interior genera
of all vertices will be zero.

In 3.3.1-3.3.4, we will assume for the sake of simplicity that there are  \it
no marked points \rm on curves. We will restore the marked points in 3.3.5.

We denote $\M(\P^n,d)$ simply by $\M$ (numbers $n$ and $d$ are supposed to be
fixed in this section).

 For a $\T$-equivariant vector bundle $\Cal E$ on the orbifold $\M^{\Gamma}$ we
denote by $[\Cal E]$ the corresponding element
of the equivariant $K$-group with rational coefficients:
$$K_{\T}^0(\M^{\Gamma})\otimes \Q\simeq
K^0(\M^{\Gamma})\otimes \Q[\T^{\vee}]\,\,.$$
In 3.3.3-3.3.4, we will denote by $[\chi]$ the element of
$K_{\T}^0(\M^{\Gamma})\otimes \Q$ corresponding to the trivial $1$-dimensional
bundle endowed with the action of $\T$ by the  (orbi-) character $\chi\in
\T^{\vee}\otimes\Q$.

 We will denote the restriction of any vector bundle $\Cal E$ on $\M$ to
$\M^{\Gamma}$ by the same symbol $\Cal E$. Often, we will denote a vector
bundle on $\M^{\Gamma}$ by its geometric fiber
 at a point $(C,f)$.
In intermediate computations in 3.3.1  we will use decompositions of fibers of
vector bundles into  formal linear combinations of some other vector spaces
arising from short exact  sequences. These auxiliary
 vector spaces will not form vector bundles, because their dimensions  will not
be  constant. Nevertheless, we will use these vector spaces  as ``vector
bundles'' putting  corresponding symbols. One can check that the final result
after all cancellations is the  class of a virtual equivariant vector bundle,
and our formal computations give the correct answer.

\subheading{3.3.1. The normal bundle}

The class of the normal bundle to $\M^{\Gamma}$ is
$$[\Cal N_{\M^{\Gamma}}]=[\Cal T_{\M}]-[\Cal T_{\M^{\Gamma}}]\,\,.$$

In 1.3.2 we computed the tangent space to $\M$:
$$[\Cal T_{\M}]=[H^0(C,f^*(\Cal T_{\P^n}))]+\dsize\sum_{y\in  C^{\alpha}\cap
C^{\beta}:\alpha\ne\beta} [T_y (C^{\alpha})\otimes T_y(C^{\beta})]\,\,+$$
$$+\left(\sum_{y\in C^{\alpha}\cap
C^{\beta}:\alpha\ne\beta}([T_y(C^{\alpha})]+[T_y(C^{\beta})])-\sum_{\alpha}[H^0(C^{\alpha}, \Cal T_{C^{\alpha}})]\right)\,\,.$$

 The first summand corresponds to infinitesimal deformations of the map $f$ of
a fixed curve $C$. The second summand corresponds to flat deformations of $C$
resolving double singular points. The third summand comes from deformations of
$C$ preserving singular points. Its first part comes from deformations of
singular points. We subtract from it the classes of the $3$-dimensional spaces
of vector fields on the irreducible components $C^{\alpha}$.

The class of the tangent space to $\M^{\Gamma}$ is
  by analogous reasons equal to
$$[\Cal T_{\M^{\Gamma}}]=\dsize\sum_{y\in  C^{\alpha}\cap
C^{\beta}:\alpha\ne\beta;\alpha, \beta\notin Edge(\Gamma)}
[T_y (C^{\alpha})\otimes T_y(C^{\beta})]\,\,+$$
$$+
\sum_{y\in C^{\alpha}\cap C^{\beta}:\alpha\ne\beta,\alpha\notin
Edge(\Gamma)}([T_y(C^{\alpha})]
\,\,-\sum_{\alpha:\alpha\notin Edge(\Gamma)}[H^0(C^{\alpha}, \Cal
T_{C^{\alpha}})]\,\,.$$

Here the first summand corresponds to resolutions of double singular points
which are intersection points of two contracted components, the second summand
comes from deformations of singular points on contracted components. Again, we
subtract classes
 of spaces of vector fields on contracted components.

Combining all the formulas above we get:
$$[\Cal N_{\M^{\Gamma}}]=[H^0(C,f^*(\Cal T_{\P^n}))]+
[\Cal N^{abs}_{\M^{\Gamma}}]\,\,,$$
where the ``absolute'' part of the normal bundle is
$$[\Cal N^{abs}_{\M^{\Gamma}}]:=\dsize\sum_{y\in
C^{\alpha}\cap C^{\beta}:\alpha\ne\beta;\alpha,\beta\in Edge(\Gamma)
} [T_y (C^{\alpha})\otimes T_y(C^{\beta})]\,\,+$$
$$+\sum_{y\in
C^{\alpha}\cap C^{\beta}:\alpha\in Edge(\Gamma),\beta\notin Edge(\Gamma)}
 [T_y (C^{\alpha})\otimes T_y(C^{\beta})]\,\,+$$
$$+\left(\sum_{y\in C^{\alpha}\cap C^{\beta}: \alpha\ne \beta,\alpha\in
Edge(\Gamma)} [T_y(C^{\alpha})]\,\,-
\sum_{\alpha:\alpha\in Edge(\Gamma)} [H^0(C^{\alpha}, \Cal
T_{C^{\alpha}})]\right)\,\,.$$

Note that the first and the third summands in the formula for $[\Cal
N^{abs}_{\M^{\Gamma}}]$ above
 are trivial vector bundles on $\M^{\Gamma}$ twisted with some characters of
the torus $\T$. Also, the term $[H^0(C,f^*(\Cal T_{\P^n}))]$ has the same
nature. Later on we will see that in all our examples all equivariant
components of the vector bundle $\Cal E$ will be trivial too ($e_i^{\Gamma,
\lambda}=0$ in the notations of 3.1).

Hence in the Bott formula applied to $\M$ we have only one term which  is not
just a multiplicative factor with values in the field of rational functions on
$Lie(\T)$. This term is
$$\dsize\sum_{y\in
C^{\alpha}\cap C^{\beta}:\alpha\in Edge(\Gamma),\beta\notin Edge(\Gamma)}
 [T_y (C^{\alpha})\otimes T_y(C^{\beta})]\,\,.$$

 We will compute the
 corresponding integrals over $\M^{\Gamma}$ in the next subsection.
  Actually, we will compute some integrals over $\M_{0,k}$ such that the
integral over
$\M^{\Gamma}$ will be equal to their product.

For an arbitrary graph, we  define a flag as an edge endowed with an
orientation (an arrow). We denote it by a pair (vertex, edge) of adjacent
cells, where the vertex is the source of the arrow on the  edge. In general,
this notation is ambiguous for graphs with simple loops. Nevertheless, we will
use it, because all graphs in our computations will be trees.

We denote natural coordinates on $Lie(\T)\simeq\C^{n+1}$ by
$\lambda_1,\dots,\lambda_{n+1}$.

\proclaim{Notation} For a flag $F=(v,\alpha)$ of $\Gamma$ we denote by $w_F$
the expression $(\lambda_{f_v}-\lambda_{f_u})/d_{\alpha}$ where $u\in
Vert(\Gamma),u\ne v$  is the second vertex of the edge $\alpha$.\endproclaim
   We consider $w_F$ as a linear function on $Lie(\T)$.
 The geometric meaning of $w_F$ is the following: it is the character of the
action of $\T$ on the tangent space to $C^{\alpha}$ at  the point $C_v\cap
C^{\alpha}$.  The flag $F=(v,\alpha)$ has a canonical dual
$\overline{F}=(u,\alpha)$ and  weights of dual flags are related by
$w_{\overline{F}}=-w_F$.

Our next goal is the computation of the contribution of $[\Cal
N^{abs}_{\M^{\Gamma}}]$ in terms of $w_F$.

 \subheading{3.3.2. Intersection theory on $\M_{0,k}$}

In this subsection
 $k$ is an arbitrary integer bigger than or equal to $3$.
 Let $w_i,\,\,i=1,\dots,k$ be a sequence of formal variables.

We compute in this subsection the following integral:
$$I(w_1,\dots,w_k):=\int\limits_{\M_{0,k}}\dsize\prod_{i=1}^k
\frac{1}{(w_i+c_1(T_{x_i}(C)))}\,\,.$$

Recall that $\M_{0,k}=\M_{0,k}(\text{point},0)$ denotes the moduli space of
stable curves $(C;x_1,\dots,x_k)$ of genus zero with marked points.

The value of the integral $I$ is a rational symmetric function in the variables
$w_i$. We can expand it as a finite Laurent series:
$$I(w_1,\dots,w_k)=\dsize\sum_{d_1,\dots,d_k\ge0:\sum d_i =k-3}
 \,\, \prod_{i=1}^k
w_i^{-d_i-1}\langle\tau_{d_1}\dots\tau_{d_k}\rangle_0\,\,,$$
where, following Witten [W], we denote by
$\langle\tau_{d_1}\dots\tau_{d_k}\rangle_0$ the rational number
$$\int\limits_{\M_{0,k}}\dsize\prod_{i=1}^k c_1(T^*_{x_i}(C))^{d_i}\,.$$

The generating function for these numbers and analogous numbers for higher
genera was predicted in [W] and computed rigorously in [K].
 The result is quite complicated. However,  for the genus zero case  the
formulas for intersection numbers are very simple. Physicists knew this formula
 already for a long time.

\proclaim{Lemma} $\langle\tau_{d_1}\dots\tau_{d_k}\rangle_0=\frac
{(k-3)!}{d_1!\dots d_k!}\,\,.$
\endproclaim

\bf{Proof:} \rm The intersection numbers for $\M_{0,k}$ are uniquely
 defined by the following properties (see [W]):
\roster
\item $\langle \tau_0\tau_0\tau_0\rangle_0=1$,
\item $\langle\tau_{d_1}\dots\tau_{d_k}\rangle_0$ is invariant under
permutations of the indices $d_*$,
\item if $d_1=0$ then
$$\langle\tau_{d_1}\dots\tau_{d_k}\rangle_0=
 \dsize\sum_{j\ge 2:d_j\ge
1}\langle\tau_{d_2}\dots\tau_{d_j-1}\dots\tau_{d_k}\rangle_0\,.$$
\endroster

One can check easily that $\frac
{(k-3)!}{d_1!\dots d_k!}$ satisfies all the conditions above. \qed

\proclaim{Corollary} $I(w_1,\dots,w_k)=\dsize\prod_{i=1}^k w_i^{-1}
\,\,\times\,\,\,\left(\sum_{i=1}^k w_i^{-1}\right)^{k-3}$ .
\endproclaim

\subheading{3.3.3. The contribution of $\Cal N^{abs}_{\M^{\Gamma}}$}

The space $\M^{\Gamma}$ is isomorphic to the product of $\M_{0,val(v)}$ over
vertices $v\in Vert (\Gamma)$
 such that their valency
 $val(v):=\#\{\text{flags }(v,\alpha)\}$ is at least $3$.
 (Recall that we omit the action of $\text{Aut}(\Gamma)$  temporarily).

The contribution of
$$\dsize\sum_{y\in
C^{\alpha}\cap C^{\beta}:\alpha\in Edge(\Gamma),\beta\notin Edge(\Gamma)}
 [T_y (C^{\alpha})\otimes T_y(C^{\beta})]$$
in the multiplicative form is equal to
$$\dsize\prod_
{v\in Vert(\Gamma),\,val(v)\ge 3}
\left(
\left(
\sum_{\text{flags }F=(v,\alpha)} w_F^{-1}\right)^{val(v)-3}
\prod_{\text{flags }F=(v,\alpha)}
w_F^{-1}\right)\,\,.$$
This formula follows from 3.3.2 and the fact that $T_y (C^{\alpha})$ is trivial
as a line bundle and that $\T$ acts trivially  on $T_y(C^{\beta})$.

The terms $$\dsize\sum_{y\in
C^{\alpha}\cap C^{\beta}:\alpha\ne \beta;\alpha,\beta\in Edge(\Gamma)}
 [T_y (C^{\alpha})\otimes T_y(C^{\beta})]$$
correspond to vertices of $\Gamma$ of valency $2$. Their contribution is
$$\dsize\prod_{v\in
Vert(\Gamma):val(v)=2}\left(w_{F_1(v)}+w_{F_2(v)}\right)^{-1}\,\,,$$
where $F_i(v),\,\,i=1,2$ are two flags containing $v$.
 Note that one can rewrite this expression as
$$\dsize\prod_{v\in Vert(\Gamma):val(v)=2}\left(\left(\sum_
{\text{flags }F=(v,\alpha)} w_F^{-1}\right)^{val(v)-3}
\prod_{\text{flags  }F=(v,\alpha)}
w_F^{-1}\right)\,\,.$$

The contribution of terms
$$-
\sum_{\alpha\in Edge(\Gamma)} [H^0(C^{\alpha}, \Cal T_{C^{\alpha}})]$$
 in the equivariant $K$-group is equal to $$-
\sum_{\alpha\in Edge(\Gamma)}\left([-w_{F(\alpha)}]+[0]+
[w_{F(\alpha)}]\right)\,\,,$$
where $F(\alpha)$ is any of two flags containing the edge $\alpha$.
 We rewrite this as
$$\dsize-\sum_{\text{flags }F}[w_F]\,\,-\sum_{\alpha\in Edge(\Gamma)
} [0]\,\,.$$

The contribution of
$$\dsize\sum_{y\in C^{\alpha}\cap C^{\beta}: \alpha\ne\beta,
\alpha\in Edge(\Gamma)} [T_y(C^\alpha)]$$
can be written as
$$\dsize\sum_{\text{flags }F=(v,\alpha):val(v)\ge 2} [w_F]\,.$$

Hence, the contribution of the last two terms in the formula for
$\Cal N^{abs}_{\M^{\Gamma}}$ is equal to
$$\dsize-\sum_{
\text{flags }F=(v,\alpha):val(v)=1} [w_F]\,\,-\sum_{\alpha\in
Edge(\Gamma)}[0]\,.$$

Let us forget for a moment about the sum of $[0]$ over edges.
Then the contribution above can be expressed in the multiplicative form as the
product of $w_F$ over flags $F=(v,\alpha)$ with $val(v)=1$. We replace $w_F$ by
$(w_F)^{-1}\left((w_F)^{-1}
\right)^{-2}$ and note that the exponent $-2$ is equal to $val(v)-3$ again.

\bf Conclusion: \rm the contribution of
$$\Cal N^{abs}_{\M^{\Gamma}}+\dsize\sum_{\alpha\in Edge(\Gamma)}[0]$$
 in the multiplicative form is equal to
$$\dsize\prod_{v\in Vert(\Gamma)}\left(\left(\sum_
{\text{flags }F=(v,\alpha)} w_F^{-1}\right)^{val(v)-3}
\prod_{\text{flags }F=(v,\alpha)}
w_F^{-1}\right)\,\,.$$

\subheading{3.3.4. The contribution of $[H^0(C,f^*(\Cal T_{\P^n}))]$}

  The space of global sections of the vector bundle $f^*(\Cal T_{\P^n})$ is
equal to the subspace of
$$\dsize\bigoplus_{\alpha\in Edge(\Gamma)}H^0(C^{\alpha},f^*(\Cal T_{\P^n}))$$
given by the condition that the values of sections at each vertex $v$  will be
the same for all edges $\alpha$ adjacent to $v$. More precisely,
 we have the following short exact sequence of equivariant vector bundles on
$\M^{\Gamma}$:
$$0@>>>H^0(C,f^*(\Cal T_{\P^n}))@>>>\dsize\bigoplus_{\alpha\in
Edge(\Gamma)}H^0(C^{\alpha},f^*(\Cal T_{\P^n}))@>>>
$$
$$@>>>\bigoplus_{v\in Vert (\Gamma)}
\left(T_{p_{f_v}}\P^n\otimes\C^{val(v)-1}\right)@>>>0$$

First, we study the contributions of  $[H^0(C^{\alpha},f^*(\Cal T_{\P^n})]$.
  The edge $\alpha$ passes through two points $p_i,p_j\in (\P^n)^{\T}$.
In some coordinate $z=(z_1:z_2)$ on $\P^1\simeq C^{\alpha}$, the map $f$
 is given by
$$ X_i(f(z))=z_1^d,\,X_j(f(z))=z_2^d,\,\,\,X_k(f(z))=0\,\text{ for }
k\ne i,j\,.$$

Here $X_k, \,k=1,n+1$ are homogeneous coordinates on $\P^n$.
 We have a short exact sequence of vector bundles on $l_{ij}$
$$0@>>>\Cal T_{l_{ij}}@>>>\Cal T_{\P^n}@>>>\Cal N_{l_{ij}}@>>>0$$
 inducing a corresponding exact sequence of vector bundles on $\M^{\Gamma}$.

One can check using this exact sequence that the following elements form a
basis of
  $H^0(C^{\alpha},f^*(\Cal T_{\P^n}))$:
\roster
\item $ z^a X_i\partial/\partial X_i,\,\,\,-d_{\alpha}\le a\le d_{\alpha}$,
\item $ z_1^a z_2^b \partial/\partial X_k,\,\,\,a+b=d_{\alpha},\,0\le
a,b,\,\,k\ne i,j$.
\endroster

Note that there is exactly one basis element ($a=0$ in the first group) on
which $T$ acts trivially. Thus, in all we have $\#Edge(\Gamma)$ terms $[0]$,
cancelling  analogous terms in 3.3.3.

The homogeneity degree of $z$ under the action of $\T$ is equal to
$w_F=(\lambda_i-\lambda_j)/d_{\alpha}$, where $F$ is a flag of $\Gamma$
containing $\alpha$ and a vertex projecting to $p_i$. The degree of the
coordinate $X_k$ is equal to $\lambda_k$.

The contribution of
$T_{p_{f_v}}\P^n\otimes\C^{val(v)-1}$ where $f_v=i$ is equal to
$$(1-val(v))\sum_{j:j\ne i} [\lambda_i-\lambda_j]\,\,,$$
because the vectors $X_i\partial/\partial X_j,\,\,j\ne i$ form a  basis of
$T_{p_i}\P^n$.

Putting all terms together we get the formula for the contribution of\linebreak
 $[H^0(C,f^*(\Cal T_{\P^n}))]-
\# Edge(\Gamma)[0]\,\,\,\,$ in the multiplicative form:
$$\prod \Sb
\alpha\in Edge(\Gamma):\\
(v_1,v_2)\text{ -  vertices of }\alpha
\endSb
\left(
\frac{
(-1)^{d_{\alpha}}\left(
\frac{d_{\alpha}}{\l_{f_{v_1}}-\l_{f_{v_2}}}\right)^{2d_{\alpha}}}
{(d_{\alpha}!)^2}
\prod_{k\ne f_{v_1},f_{v_2}}
\prod_{a,b\ge 0:a+b=d_{\alpha}}\frac{1}{\frac{a}{d_\alpha}\l_{f_{v_1}}+
\frac{b}{d_\alpha}\l_{f_{v_2}}-\l_k}
\right)\times$$
$$\times\,\,
\prod_{v\in Vert(\Gamma)} \left(\prod_{j:j\ne
f_v}(\lambda_{f_v}-\lambda_j)\right)^{val(v)-1}\,.$$

\subheading{3.3.5. Marked points}

The only result of the introduction  of marked points is that one has to
replace the exponent $\,(val(v)-3)\,$ in the last  formula in 3.3.3
 by $\,(val(v)-3+\# S_v)$.
  We leave all checking to the reader.

It is reasonable to change a little bit the graphs associated with
 connected components of the fixed point set.
  Namely, we replace any graph $\Gamma$ with specifications as in 3.2 by a  new
graph $\widehat\Gamma$. The vertices of $\widehat\Gamma$
 are the vertices of $\Gamma$ together with a $k$-element set
$T=\,Tail\,(\widehat\Gamma)$ (tails  of $\widehat\Gamma$). The elements of $T$
are numbered from $1$ to $k$, where the tail $t$ has number $i(t)$. The edges
of $\widehat\Gamma$ are the edges of $\Gamma$ together
 with one edge $\alpha_t$ for each tail $t\in T$ connecting $t$ with the unique
vertex $v$ of $\Gamma$ such that $i(t)\in S_v$. Also we define
 $f_t$ to be equal to $f_v$ for $i(t)\in S_v$.
 We pose $d_{\alpha_t}$ to be equal to $0$ for all $t\in T$. Then,  for any
flag $F$ of $\widehat\Gamma$ containing a tail as an edge,
  one has formally $w_F^{-1}=0$.

In the sequel we will denote $\M^{\Gamma}$ by $\M^{\widehat\Gamma}$.

\subheading{3.3.6. The contributions of the vector bundles in the examples}

As we already mentioned in 3.3.1, in all three examples from  section 2 the
vector bundles arising on $\M_{0,*}(\P^n,d)^{\Gamma}$
 split into direct sums of trivial bundles twisted with characters of $\T$.
Thus, their contributions will be  rational functions on $Lie(\T)$ depending on
$\Gamma$. The computation of these contributions is easier than that of the
previous ones.

In  example 2.1 the contribution of line bundles $\Cal O(1)_i$ is
$$\prod_{v\in Vert(\Gamma)}(\lambda_{f_v})^{2\# S_v}\,\,,$$
or, using the modified graph $\widehat\Gamma$,
$$\prod_{t\in \,Tail\,(\Gamma)}\lambda^2_{f_t}\,\,.$$

In  example 2.2 the contribution of $\Cal E_d$ is
$$\prod_{\alpha\in Edge(\Gamma):(v_1,v_2)\text{ -  vertices of
}\alpha}\left(\prod_{a,b\ge 0:
a+b=5d_{\alpha}}\frac{a\lambda_{f_{v_1}}+b\lambda_{f_{v_2}}}{d_{\alpha}}\right)\times
\prod_{v\in Vert(\Gamma)}(5\lambda_{f_v})^{1-val(v)}\,\,.$$
Here we use the short exact sequence (omitting zeroes from the left and the
right):
$$H^0(C,f^*(\Cal O (5)))@>>>\dsize\bigoplus_{\alpha\in Edge(\Gamma)}
H^0(C^{\alpha},f^*(\Cal O (5)))@>>>\bigoplus_{v\in Vert (\Gamma)}
\left(\Cal O(5)_{p_{f_v}})\otimes \C^{val(v)-1}\right)\,\,\,.$$

In  example 2.3 the contribution of $\Cal F_d$ is
$$\left[\prod_{\alpha\in Edge (\Gamma): (v1,v2)\text{ - vertices of
}\alpha}\left(
\prod_{a,b<0:a+b=-d_{\alpha}}
\frac{a\lambda_{f_{v_1}}+b\lambda_{f_{v_2}}}
{d_{\alpha}}
\right)\times\prod_{v\in Vert (\Gamma)}
(-\lambda_{f_v})^{val(v)-1}\right]^2\,.$$
 It follows from the short exact sequence
$$\bigoplus_{v\in Vert (\Gamma)}
\left(\Cal O(-1)_{p_{f_v}})\otimes \C^{val(v)-1}\right)@>>>H^1(C,f^*(\Cal O
(-1)))@>>>\dsize\bigoplus_{\alpha\in Edge(\Gamma)}
H^1(C^{\alpha},f^*(\Cal O (-1)))
\,\,\,.$$

\subheading{3.4. The final sum}

In each example the integral over the corresponding moduli space of stable maps
 is equal to the sum over equivalence classes of appropriate graphs
 of
$$\frac{1}{\# Aut(\widehat\Gamma)}(\text{ formula from 3.3.3})(\text{ formula
from 3.3.4})(\text{ formula from 3.3.6})\,\,.$$
The last formula from 3.3.3 is corrected in the first example according to
3.3.5.

\head 4. Critical values.\endhead

\subheading{4.1. Feynman rules and summation over trees}

Here we will describe a general formula, well known in physics and
combinatorics, which gives values of certain infinite sums over trees. For
additional information on summation over trees and graphs one can look at
chapter $7$ in [ID].

The initial data consist of a finite or countable set of indices $ A$, a
symmetric non-degenerate matrix $g=(g^{ab}),\,\,g^{ab}=g^{ba},\,\,a,b\in A$,
 and an infinite sequence of symmetric tensors with lower indices:
$$C_{a_1\dots a_k},\,\,\,a_i\in A,\,\,\,k\ge 0\,\,.$$
  Here the coefficients of the tensors $g^*,C_*$ are complex numbers or
elements of a topological field of characteristic zero (for example, a field of
formal power series in auxiliary variables with coefficients in $\C$). We
assume that all infinite series  appearing  later are convergent in an
appropriate topology.

These data define a function on the set of equivalence classes of finite
graphs.
 Let $\Gamma$ be a graph and $Flags(\Gamma)$ be the set of its flags. The
weight of the graph is defined as
$$w(\Gamma):=\sum\Sb\text{maps }\\f:Flags(\Gamma)@>>>A\endSb\left(
\prod \Sb
\alpha\in Edge(\Gamma)\\
F_1,F_2\text{ - flags}\\\text{containing }\alpha
\endSb
g^{f(F_1)f(F_2)}
\prod\Sb v\in Vert(\Gamma)\\F_1,\dots,F_k\text{
 - flags}\\\text{ containing }v\endSb
C_{f(F_1)\dots f(F_k)}\right)\,\,.$$

Note that in all our examples one can choose an appropriate set of indices and
tensors $g,C$ such that the sum of contributions of connected components
 of $\M^{\widehat{\Cal \Gamma}}$ corresponding to any abstract tree (without
specifications $d_{\alpha},f_v, S_v$) will be equal to the weight
 of this tree. We will show in 4.1.1 how to choose $A,g,$ and $C$
in all our examples.

Define  the ``tree-level partition function'' (just an element of the ground
field) by the formula
$$Z^{tree}:=\sum \Sb \Gamma:\text{ equivalence classes of}
\\ \text{finite nonempty trees }\Gamma\endSb \frac{1}{\#
Aut(\Gamma)}\,w(\Gamma)\,\,.$$

Let us introduce auxiliary formal variables $\phi_a, \,a\in A$ and a series
$$S(\phi_*):=-\sum_{a,b\in A} \frac{g_{ab}\phi_a\phi_b}{2}
+\sum_{k\ge 0}\frac{1}{k!}\sum_{a_1,\dots,a_k\in A}C_{a_1\dots a_k}
\phi_{a_1}\dots\phi_{a_k}\,\,.$$
Here $g_{ab}$ denote the matrix coefficients of the inverse matrix
$(g)^{-1}$:
$$(g_{ab})=(g^{ab})^{-1}\,\,.$$

Later it will be convenient to refer to the first summand in the formula for
$S$ as to the ``kinetic'' part and to the second summand
 as to the ``potential'' part of the ``action'' functional $S$, in analogy with
 classical mechanics.

\proclaim{Formula} We have $Z^{tree}=Crit\,S(\phi_*)$, where the r.h.s. denotes
the critical value of
 the function $S$.
\endproclaim

This formula follows from a more general formula
$$ log\left((det(2\pi \hbar g))^{-1/2}\int
e^{\frac{S(\phi_*)}{\hbar}}\prod_{a\in A}d\phi_a\right)=\sum\Sb \Gamma:\text{
equivalence classes of}
\\ \text{connected nonempty graphs }\Gamma\endSb
\frac{\hbar^{-\chi(\Gamma)}}{\# Aut(\Gamma)}\,w(\Gamma)\,\,,$$
 where both sides of the formula are considered as formal power series
expansions at $\hbar@>>>0$.

The last formula is the usual expansion of integrals over Feynman diagrams.

The argument above is valid only in the case of coefficients with values in
$\C$ when $A$ is finite, $g^*$ is real-valued and positive-definite,
 $C_*$ are sufficiently small
 and the integral above is convergent.

A direct proof of the formula $Z^{tree}=Crit\,S(\phi_*)$ can be obtained by a
formal inversion of the map $(\phi_*)\mapsto dS_{|(\phi_*)}$ and evaluating $S$
at the critical point.

\subheading{4.1.1. The action functionals in the three examples}

Let us first describe the situation without marked points on the curves (and
without tails on the graphs).

We denote the  variables by $\phi_{ij,d}$ where $i\ne j,\,\,\,i,j\in
\{1,\dots,n+1\},\,\,\,d\ge 1$. Hence,
 the set of indices is
$$A= \left(\{1,\dots,n+1\}^2\setminus\,diagonal\,\right)\times\bold N\,\,\,.$$

For any graph $\Gamma$, we put on each flag $F=(v,\alpha)$
 the composite index $ij,d\,\,$, where $d=d_{\alpha},\,i=f_v,\,j=f_u$ (here, as
usual, $u$ denotes the second vertex of $\alpha$).

The potential part of the action is standard:
$$S^{pot}(\phi_*)=\sum_{i:1\le i\le n+1}\mu_i c_i\sum_{k\ge 1}
\frac{1}{k!}\sum\Sb j_1,\dots,j_k:j_*\ne i
\\ d_1,\dots ,d_k: d_*\ge 1
\endSb
(v_{i j_1,d_1}+\dots+v_{i j_k d_k})^{k-3}
\phi_{i j_1,d_1}\dots \phi_{i j_k d_k}\,\,,$$
where $v_{ij,d}:=d/(\lambda_i-\lambda_j)$ is equal to $w_F^{-1}$
 for corresponding flags $F$
 and
$$\mu_i=\prod_{j:j\ne i}(\lambda_i-\lambda_j)\,\,,$$
and where $c_i$ is some constant depending on the situation.
 The potential part consists of the contribution of $\Cal
N^{abs}_{\M^{\Gamma}}$ divided by the product of $w_F^{-1}$
 and multiplied by one factor $\mu_i$ in the contribution of $[H^0(C,f^*(\Cal
T_{\P^n}))]$, and $c_i$ coming from the vector bundle on $\M$.
We will find another formula for   $S^{pot}(\phi_*)$ in the next subsection.

The coefficients  $g^{ij,d;i'j',d'}$ will be non-zero only for
  $i=j',\,j=i',\,d=d'$. This guarantees that graphs with indices on flags which
have non-zero weight will be in one-to-one correspondence with graphs with
specifications of the type introduced in 3.2.
We will denote $g^{ij,d;ji,d}$ simply by $g^{ij,d}$. The inversion of the
matrix $g$ is reduced to the inversions of the numbers  $g^{ij,d}$.

The contribution of $[H^0(C,f^*(\Cal T_{\P^n}))]$ in
 $g^{ij,d}$ (with removed factor $\mu_i$ and added factors $w_F^{-1}$) is
equal to
$$-w_F^{2d-2}(d!)^2\mu_i^{-1}\mu_j^{-1}\,\prod_{k\ne i,j} \,\,\,\prod_{a,b\ge
0:a+b=d}\left(
\frac{a\lambda_i+b\lambda_j}{d}-\lambda_k\right)^2 \,\,.$$

The contribution of  $\Cal E_d$ in the example
 2.2 is equal to
 $$(25\lambda_i\lambda_j)^{-1}\,\,\,\prod_{a,b\ge 0:a+b=5d}
\left(\frac{a\lambda_i+b\lambda_j}{d}\right)\,\, \,.$$
The constant $c_i$ in the potential part is $5\lambda_i$.

In the example 2.3, the contribution of $\Cal F_d$ is equal to
$$\lambda_i\lambda_j\,\,\,\prod_{a,b< 0:a+b=1-d}
\left(\frac{a\lambda_i+b\lambda_j}{d}\right) \,\,\,.$$
The constant $c_i$ in the potential part is $(-\lambda_i)^{-1}$.

In both examples 2.2 and 2.3, we multiply $g^{ij,d}$ by $z^d$, where
 $z$ is a new formal variable. Thus the resulting critical value will be a
series in $z$ with coefficients equal to $N_d$ and $M_d$ respectively.

In the example 2.1, we add variables $\tilde \phi_i\,,\,\,\,1\le i\le n+1=3$.
The matrix $g$ is the same as above for indices from the set $A$ (no
contributions from vector bundles), and
$$g^{i;ij,d}=0,\,\,\,g^{i:j}=\delta_{ij} \,\lambda _i^2\,.$$
Here $\delta_{ij}$ is the Kronecker symbol.
 The potential part is equal to
$$S^{pot}(\phi_*,\tilde\phi_*)=\sum_{i:1\le i\le n+1}\mu_i \sum_{k\ge 1}
\frac{1}{k!}\sum_{l\ge 0}\frac{1}{l!}$$
$$\left(\sum\Sb j_1,\dots,j_k:j_*\ne i
\\ d_1,\dots ,d_k: d_*\ge 1
\endSb
\,\,\sum_{\tilde j_1,\dots,\tilde j_l}
(v_{i j_1,d_1}+\dots+v_{i j_k d_k})^{k+l-3}\,
(\phi_{i j_1,d_1}\dots \phi_{i j_k d_k})
(\tilde\phi_{\tilde j_1}\dots\tilde\phi_{\tilde j_l})\right)\,\,.$$
 As before, we multiply the matrix coeficients of $g$ by extra variables:
 $$g^{ij,d}\,\mapsto \, z_1^d\,g^{ij,d},\,\,\,g^{i;i}\,\mapsto\,
z_2\,g^{i,i}\,\,.$$
The exponents of the variables $z_1,z_2$ count the total degree of the curves
and the number of marked points respectively. Then in the resulting sum over
trees considered as a series in $z_1,z_2$ we have to extract monomials of the
form $z_1^d z_2^{3d-1}$. It can be done by a contour integration.

\subheading{4.2. The potential part of the action functional}

$S^{pot}(\phi_*)$ comes essentially  from  the intersection numbers of
$\M_{0,k}$.

In the sequel of this subsection we use some auxiliary set of indices $I$ and
two sequences of variables $v_i,\,\phi_i,\,\,\,i \in I$. Later we will set
$v_i:=w_*^{-1}$. In the notations of the previous subsection, we consider only
a summand of $S^{pot}$ corresponding to a fixed point $p_*\in \P^n$ and omit
multiples $\mu_*,c_*$.

Define a function $S$ by the formula
$$S(v_*,\phi_*):=\sum_{k\ge 1}\frac{1}{k!}\sum_{i_1,\dots,i_k \in I}
 (v_{i_1}+\dots+v_{i_k})^{k-3}
\phi_{i_1}\dots \phi_{i_k}\,\,.$$

\proclaim{Theorem} We have $S(v_*,\phi_*)=Crit\,B(\xi)$, where
$$B(\xi)=\frac{\xi^3}{6}+\frac{1}{2}\sum_{i,j\in I}
\phi_i \phi_j\,\frac{ exp\,(\xi v_i+\xi v_j)}{v_i+v_j}+
\sum_{i\in I} \phi_i\, \frac{exp\,(\xi v_i)}{v_i^2}-
\sum_{i\in I}\phi_i \,\frac{ \xi \,\,exp\,(\xi v_i)}{v_i}\,\,.$$
\endproclaim

Note that in the final form we will have $n+1$ distinct variables
 $\xi _i$ corresponding to different parts of $S^{pot}$.

The rest of this subsection will be devoted to the proof of the theorem.

First of all,  we dismantle the definition of $S$ into simple pieces, doing the
opposite to what was done in 3.3.3.
 Maybe it is not the most economical way to prove our formula.

Denote by $S_k$ the $k$-th summand in the definition of $S$.
For $k\ge 3$, we have:
$$S_k(v_*,\phi_*)=\frac{1}{k!}\sum_{i_1,\dots,i_k \in I}
 (v_{i_1}+\dots+v_{i_k})^{k-3}
\phi_{i_1}\dots \phi_{i_k}=$$
$$=\frac{(k-3)!}{k!}\,\sum_{i_1,\dots,i_k \in I}\,\,\,\sum_{d_1,\dots,d_k\ge
0:\sum d_j=k-3} \,\,\,\prod_{j=1}^k \frac{\phi_{i_j}v_{i_j}^{d_j}}{d_j!}\,\,.$$

Let us introduce more notations:
$$\Cal Z_d:=\sum_{i\in I} \phi_i \frac {v_i^d}{d!},\,\,\,\Cal
Z:=\sum_{d=0}^{\infty} \Cal Z_d \xi ^{d-1}=\xi ^{-1}\sum_{i\in I}\phi_i
\,exp\,(\xi v_i)\,\,\,.$$

We can rewrite the formula for $S_k$ above as
$$S_k(v_*,\phi_*)=
\frac{(k-3)!}{k!}\,\,\hbox{Coeff}_{\xi ^{-3}}\left(\sum_{d_1,\dots,d_k\ge 0}
(\Cal Z_{d_1} \xi ^{d_1-1})\dots(\Cal Z_{d_k} \xi ^{d_k-1})\right)=$$
$$=\frac{(k-3)!}{k!}\,\,\hbox{Coeff}_{\xi ^{-3}}\Cal Z^k \,\,.$$

Thus,
$$\sum_{k\ge 3} S_k=\,\hbox{Coeff}_{\xi ^{-3}}
\left(
\sum_{k\ge 3}\frac{(k-3)!}{k!}\Cal
Z^k\right)=\,\hbox{Coeff}_{\xi ^{-3}}\left(\widetilde
{\Psi}(\Cal Z)\right)\,,$$
where
$$\widetilde
{\Psi}(\Cal Z):=\frac{{\Cal Z}^3}{1\cdot2\cdot3}+\frac{{\Cal
Z}^4}{2\cdot3\cdot4}+\dots=\frac{(1-\Cal Z)^2}{2}log\frac{1}{1-\Cal
Z}+\frac{3}{4}\Cal Z^2-\frac {\Cal Z}{2}\,\,.$$

We can replace  $\widetilde
{\Psi}(\Cal Z)$ by ${\Psi}(\Cal Z)=\frac{(1-\Cal Z)^2}{2}log\frac{1}{1-\Cal
Z}$, because we take the coefficient of the monomial $\xi ^{-3}$ and $\Cal Z$
has a pole of the first order at $\xi =0$.

$$\hbox{Coeff}_{\xi ^{-3}}\left(
{\Psi}(\Cal Z)\right)=\frac{1}{2\pi i}\oint\limits_{|\xi |=1}\frac{(1-\Cal
Z)^2}{2}log\frac{1}{1-\Cal Z} \,\,\xi ^2 d\xi =$$
$$= \frac{1}{2\pi i}\oint\limits_{|\xi |=1} A'(\xi )\,log\,\frac{1}{1-\Cal
Z(\xi )}d\xi \,,$$
where the regular function $A(\xi )$ is defined by the conditions
$$A'(\xi )=\frac{(\xi -\xi \Cal Z(\xi ))^2}{2}\,,\,\,\,\,A(0)=0\,\,.$$
Now we can integrate by parts:
$$\hbox{Coeff}_{\xi ^{-3}}\left(
{\Psi}(\Cal Z)\right)=-\frac{1}{2\pi i}\oint\limits_{|\xi |=1}
 A(\xi )\left(log\frac{1}{1-\Cal Z(\xi )}\right)'d\xi =$$
$$=-\frac{1}{2\pi i}\oint\limits_{|\xi |=1}
 A(\xi )\frac{\Cal Z'(\xi )}{1-\Cal Z(\xi )}d\xi =res_{\xi
_0}\left(A(\xi)\,d\,log(1-\Cal Z(\xi))\right)=A(\xi _0)\,.$$
 Here $\xi _0$ is the root of the equation $\Cal Z(\xi )=1$. Note that
 by the definition of $A(\xi )$ its derivative at $\xi _0$ vanishes.
 Hence,
$$\hbox{Coeff}_{\xi ^{-3}}\left(
{\Psi}(\Cal Z)\right)=\,Crit\,\,A(\xi )\,.$$

We can compute $A(\xi )$ explicitly:
$$A'(\xi )=\frac
{(\xi -\xi \Cal Z(\xi ))^2}{2}=\frac{\xi ^2}{2}+\frac{1}{2}\left(\sum_{i\in
I}\phi_i \,exp\,(\xi v_i)\right)^2-\xi \sum_{i\in I}\phi_i \,exp\,(\xi
v_i)\,,$$
$$A(\xi )=\frac{\xi ^3}{6}+\frac{1}{2}\sum_{i,j\in I}
\phi_i \phi_j\,\frac{ exp\,(\xi v_i+\xi v_j)-1}{v_i+v_j}+$$
$$+\sum_{i\in I}
 \phi_i\frac{exp\,(\xi v_i)-1}{v_i^2}-\sum_{i\in I}\phi_i\frac{\xi \,exp\,(\xi
v_i)}{v_i}\,.$$

After adding to the formula above two terms $S_1(v_*,\phi_*)$ and
 $S_2(v_*,\phi_*)$ which do not depend on $\xi$, we obtain $B(\xi )$. \qed
\subheading{4.2.1. The effect of marked points}

We need a generalization
  of the  previous computations to the case when some $v_i$ are equal to zero
(for curves with marked points).

Now we have two groups of indices $I, \,J$, and variables
  $v_i,\,\phi_i,\,\,i\in I$, and $\tilde \phi_j,\,\,j\in J$.
 The potential $S(v_*,\phi_*,\tilde \phi_j)$ is defined by the formula
$$S:=\sum_{k\ge 1}\frac{1}{k!}\sum_{l\ge 0}\frac{1}{l!}\sum_{i_1,\dots,i_k \in
I}\,\sum_{j_1,\dots,j_l\in J}
 (v_{i_1}+\dots+v_{i_k})^{k+l-3}
(\phi_{i_1}\dots \phi_{i_k})(\tilde\phi_{j_1}\dots\tilde\phi_{j_l})\,\,.$$

One can  rewrite it as
$$\sum_{k\ge 1}\frac{1}{k!}\sum_{i_1,\dots,i_k \in I}
 (v_{i_1}+\dots+v_{i_k})^{k-3}
\phi_{i_1}\dots \phi_{i_k}\,exp\left(\widetilde\Phi
(v_{i_1}+\dots+v_{i_k})\right)
\,\,,$$
where $\widetilde\Phi=\dsize\sum_{j\in J}\tilde \phi_j$.
Thus, we can use the previous formulas from 4.2 with $\phi_i$ replaced
 by $\phi_i\,exp\,(\widetilde\Phi\,v_i)$.

\subheading{4.3. The structure of the resulting functional}

Note that we have the following general scheme in all examples:
  the sum over trees is equal to the extremal values of a function in
infinitely many variables $\phi_*$ and finitely many variables $\xi _*$ and
$\tilde \phi_*$ (the variables $\lambda$ are considered as constant and the
result should not depend on them). The resulting functional is \it quadratic
\rm in $\phi$. Thus, we can, in principle, find its extremal
 value by solving a system of linear equations. This system is infinite and it
is not easy to solve it. At the moment we do not know how to proceed.



\subheading{4.4. Examples}

We will not write the formula for the example 2.1.

It is convenient to rescale the variables $\phi_*$ as
 $$\phi_{ij,d}\mapsto \phi_{ij,d}\,exp\,(\phi_i v_{ij,d})\,.
$$

For 2.2 (curves on quintics) the functional is
$$\frac{1}{2}\sum\Sb i,j;d\\i\ne j\endSb \frac
{d^3 (\l_i-\l_j)^2\dsize\prod_{a+b=d:a,b\ge 1}\dsize\prod_{k=1}^5
(a\l_i+b\l_j-d\l_k)}
{\dsize\prod_{a+b=5d:a,b\ge 1}(a\l_i+b\l_j)}
\,exp\left(-td-\frac{\xi_i-\xi_j}{\l_i-\l_j}\right)\phi_{ij,d}\phi_{ji,d}\,+$$
$$+\frac{1}{2}\sum\Sb i,j,d,j',d'\\ j,j'\ne i\endSb \nu_i
\frac
{
\phi_{ij,d}\phi_{ij',d'}
}
{
\frac{d}{\l_i-\l_j}+\frac{d'}
{\l_i-\l_{j'}}
}\,\,-$$
$$-\sum\Sb i,j,d\\i\ne j\endSb\nu_i\frac{\l_i-\l_j}{d}\xi_i\phi_{ij,d}+
\sum\Sb i,j,d\\i\ne
j\endSb\nu_i\frac{(\l_i-\l_j)^2}{d^2}\phi_{ij,d}+\sum_{i=1}^5\nu_i\frac{\xi_i^3}{6}\,\,.$$
Here $\nu_i$ denotes $\frac{5\l_i}{\prod_{j:j\ne i} (\l_i-\l_j)}$.
The extremum is taken over the variables $\phi_*$ and $\xi_*$. The result does
not depend on $\l_*$ and is equal (conjecturally) to the function $F(t)$ from
2.2 minus $5t^3/6$.

 If we specify $\l$ to be equal to roots of $1$:
$\,\,\,\l_j:=exp(2\pi \sqrt{-1} j/5)\,\,,$
then by homogeneity at the extremal point we have $\xi_j=\Xi \l_j$ for some
$\Xi=\Xi(t)$. The last summand in the formula for the functional is equal to
$5\Xi^3/6$. Note that $\Xi$ is in a sense opposite to $t$
 because in the first term we have the exponent
  $$exp\,\left( -(t+\Xi)d\right)\,\,.$$

In general we expect that the cubic term from the main formula of 4.2
  corresponds to the contribution of maps of degree zero in the potential (see
[KM]).

 For 2.3 (multiple coverings), we will write  only a formula for the functional
 with fixed $\l$-s: $\lambda_1=-\lambda_2=1$. After some simplifications and
changings of notations we get the following formula:
$$\sum_{k,l}\frac{\phi_k\phi_l}{k+l}-\sum_k\frac{2^{4k}}{k}{\binom {2k} k
}^{-2}
\,exp\,\left(-k(t+X)\right)\,\phi_k^2-\frac{X}{2}\sum_k \frac{\phi_k}{k}
+\sum_k\frac{\phi_k}{k^2}+\frac{X^3}{6}\,\,.$$
 The indices $k,l$ take values in the set $\{1,3,5,\dots\}$.  Recently Yu.
Manin (see [M]) noticed that if one of the variables $\l_*$ vanishes then the
sum over trees simplifies drastically because the contribution of the most part
of trees dissapear. The result of the compuation   is equal to $Li_3(e^t)$.

\head 5. Generalizations.
\endhead

\subheading{5.1. Higher genera}

We do not know at the moment how to treat higher genus curves
in the same way as we treat rational curves. One basic problem is that the
moduli space of stable maps is almost never smooth for higher genus.
  Maybe, this is not a serious obstruction if one adopts the general philosophy
of hidden smoothness presented in 1.4, and one can apply
 Bott's formula ignoring singularities.
 If all this will work, we will use intersection numbers
 on $\M_{g,k}$ including the numbers computed in [K] and, maybe, something
else.

We hope, that the generating function over all genera (``string partition
function'') will be equal to a sum over graphs and that it will reduce finally
to a Feynman integral with auxiliary matrices (as in [K]). Also, we  hope that
the trick from 4.2 will work in the quantum case too, reducing the final
integral to a finite-dimensional one via an elimination of free fields (i.e.,
gaussian integrals).

The simplest case from which it is reasonable to start is the case of the
projective plane. How many curves of genus $g$ and degree $d$ in $\P^2$ pass
through generic $3d-1+g$ points? There was a recursion formula proposed by
Z.~Ran in [R].
 Unfortunately, it is not correct, because of misprints or mathematical errors.
 At least Yu.~Manin and the author were not able to get the number $12$ of
 plane cubics passing through generic $8$ points using Ran's formulas.

\subheading{5.2.   Flag spaces and toric varieties}

All of our computational scheme works well for any generalized flag space
$G/P$, where $G$ is a semi-simple algebraic group and $P$ is a parabolic
subgroup.

All we need is that the moduli spaces of genus zero stable maps are smooth and
that the Cartan subgroup $T$ has isolated fixed points and isolated
 $1$-dimensional orbits on $G/P$.

The first problem  is to compute genus zero Gromov-Witten invariants on flag
varieties. As was explained in [KM], heuristic arguments of [GK] do not give
the whole information encoded in the potential.

Moduli spaces of genus zero stable maps to a toric variety
 are not smooth in general. Nevertheless, it is possible that the  Bott formula
can be modified and applied again.

\subheading{5.3. Complete intersections}

We can treat any smooth complete intersection of hypersurfaces in projective
space in the same manner as quintic $3$-folds.
  In generalizations to other varieties endowed with a torus action,  we should
consider equivariant vector bundles generated by global sections. If we have
realized $V$ as the zero set of a section of such a vector bundle transversal
to the zero section, then all the machinery applies.

\subheading{5.4. Families}

Let us consider counting problems of genus zero curves on $K3$-surfaces. The
first impression is that it should be trivial, because there are no non-trivial
curves on non-algebraic K\"ahler surfaces and Gromov-Witten classes are
invariant under deformations.

Let us consider now a $1$-parameter holomorphic family of $K3$-surfaces $S_t$
 such that $S_0$ is algebraic and, for almost all $t$, $S_t$ is not algebraic.
   The union $\Cal S:=\bigcup_t S_t$ is a non-compact $3$-dimensional complex
variety with  trivial first Chern class.
  Hence, we expect that  compact rational curves on a generic small
almost-complex  perturbation of $\Cal S$ are isolated and there will be
finitely many of them
 sitting on $S_0$ in the limit as the perturbation tends to zero.
 D.~Morrison (private communication) proposed to consider a particular
$1$-parameter deformation of $K3$-surfaces, namely, the twistor family
 of K\"ahler structures with a fixed Ricci-flat metric.

It seems that considering Gromov-Witten invariants of total spaces of families
is reasonable only for genus zero curves, otherwise parasitic contributions to
the virtual tangent bundle appear.

For $V$ being a generic quartic surface in $\P^3$ the Picard group
$Pic(V)\simeq \Z$ is generated by the plane section. Thus, degrees of curves on
$V$ are divisible by $4$. The dimension of the space of curves of degree $4d$
on $V$ is equal to $2d^2+1$ and the genus of the generic  curve of degree $4d$
is also equal to $2d^2+1$. Hence, we expect
 a finite number of rational curves of degree $4d$ with
  $2d^2+1$ nodes.
We believe that these numbers fit into the picture above, because a generic
quartic has a canonical $1$-st order non-algebraic deformation.
  Unfortunately, we were not able to define the numbers of rational curves on
quartics following the pattern of section 2. Presumably,
 there should be a Mirror relation between these numbers and a variation of
Hodge structures with one of the periods equal to
$$I(z)=\sum_{n\ge 0} \frac{(4n)!}{(n!)^4}z^n\,\,.$$
\newline
\newline
\it Acknowledgements\rm. The author is grateful to Yu.~Manin and F.~Cukierman
for useful comments, and to the Max-Planck-Institut
 f\"ur Mathematik in Bonn for the hospitality and stimulating atmosphere.


\Refs

\widestnumber\key{GMM}


\ref \key AM\by P. S. Aspinwall, D. R. Morrison \paper
 Topological field theory and rational curves \jour Commun. Math. Phys. \pages
245--262\yr 1993 \vol 151\endref


\ref \key BFM\by P. Baum, W. Fulton, R. MacPherson\paper
 Riemann-Roch for singular varieties\pages 101--145\yr 1975\jour Publ. Math. I.
H. E. S. \vol 45\endref


\ref \key B\by R. Bott\paper A residue formula for holomorphic vector
fields\jour Jour. Diff. Geom.\yr 1967 \vol 1 \pages 311-330\endref


\ref \key DM\by P. Deligne, D. Mumford\paper The irreducibility of the space of
curves of given genus\pages 75-110\yr 1969\jour Publ. Math. I. H. E. S. \vol
36\endref

\ref \key GK\by A. Givental, B. Kim \paper Quantum cohomology of flag manifolds
and Toda lattices\yr 1993 \jour preprint\endref

\ref \key ID\by  C. Itzykson, J.-M. Drouffe\book Statistical field theory \yr
1989\publ Cambridge University Press\endref


\ref \key K\by M. Kontsevich\paper Intersection theory on the moduli space of
curves and the matrix Airy function \pages 1--23\vol 147 \yr 1992 \jour Commun.
Math. Phys.\endref

\ref \key KM\by M. Kontsevich, Yu. Manin\paper Gromov-Witten classes, quantum
cohomology and enumerative geometry\yr 1994\jour MPI preprint and
hep-th/9402147\endref


\ref \key P\by P. Pansu \paper Chapter VIII, Compactness\inbook Holomorphic
curves in symplectic geometry, eds. M. Audin, J. Lafontaine, Progress in
Mathematics\vol 117\publ Birkh\"auser\yr 1994\endref

\ref\key R\by Z. Ran\paper Enumerative geometry of singular plane curves
\jour Invent. Math. \yr 1989 \pages 447--465\vol 97\endref

\ref \key RT\by Y. Ruan, G. Tian\paper Mathematical theory of quantum
cohomology \jour preprint \yr 1993\endref


\ref \key W\by E. Witten \paper Two-dimensional gravity and intersection theory
on moduli space \yr 1991\vol 1 \pages 243--310 \jour Surveys in Diff.
Geom.\endref

\ref \key Y\by S. T. Yau, ed. \book Essays on Mirror Manifolds\yr 1992  \publ
International  Press Co. , Hong Kong\endref

\endRefs
\enddocument